# Ultrafast insulator-to-metal transition in VO$_2$ nanostructures assisted by picosecond strain pulses

Ia. A. Mogunov,[1,*] F. Fernández,[2] S. Lysenko,[2] A. J. Kent,[3] A. V. Scherbakov,[1,4] A. M. Kalashnikova,[1] and A. V. Akimov[3]

[1] *Ioffe Institute, 194021 St. Petersburg, Russia*

[2] *Department of Physics, University of Puerto Rico, Mayaguez, 00681 Puerto Rico, USA*

[3] *School of Physics and Astronomy, University of Nottingham, Nottingham NG7 2RD, UK*

[4] *Experimentelle Physik 2, Technische Universität Dortmund, D-44227 Dortmund, Germany*

[*]corresponding author: mogunov@mail.ioffe.ru

Abstract

Strain engineering is a powerful technology which exploits stationary external or internal stress of specific spatial distribution for controlling the fundamental properties of condensed materials and nanostructures. This advanced technique modulates in space the carrier density and mobility, the optical absorption and, in strongly correlated systems, the phase, e.g. insulator/metal or ferromagnetic/paramagnetic. However, while successfully accessing nanometer length scale, strain engineering is yet to be brought down to ultrafast time scales allowing strain-assisted control of state of matter at THz frequencies. In our work we demonstrate a control of an optically-driven insulator-to-metal phase transition by a picosecond strain pulse, which paves a way to ultrafast strain engineering in nanostructures with phase transitions. This is realized by simultaneous excitation of VO$_2$ nanohillocks by a 170-fs laser and picosecond strain pulses finely timed with each other. By monitoring the transient optical reflectivity of the VO$_2$, we show that strain pulses, depending on the sign of the strain at the moment of optical excitation, increase or decrease the fraction of VO$_2$ which undergoes an ultrafast phase transition. Transient strain of moderate amplitude ~0.1% applied during ultrafast photo-induced non-thermal transition changes the fraction of VO$_2$ in the laser-induced phase by ~1%. By contrast, if applied after the photo-excitation when the phase transformations of the material are governed by thermal processes, transient strain of the same amplitude produces no measurable effect on the phase state.

# I. Introduction

During the last decades, strain engineering has developed into a powerful technology to control the electron density and mobility in semiconductors. Strain engineering is used particularly for fabrication of silicon based integrated circuits in microprocessors [1] and optical devices [2, 3]. Nowadays the interest in emerging quantum technologies and further miniaturization of electronic and optical devices has turned strain engineering towards nanoobjects like two-dimensional layers (for reviews see Ref. [4, 5]), quantum dots [6, 7] and nanotubes [8]. Strain engineering utilizes stationary spatial strain distributions for band gap engineering [9, 10] achieving high pseudo-magnetic fields [11] and anisotropic current channels [12, 13]. Strain engineering has also been proposed for magnetic phase separation [14] and signal processing [15].

It is appealing to extend strain engineering to the ultrafast temporal scale and control the electrons, lattice, and spins both in space and time. It has already been shown that picosecond strain pulses can be successfully used for ultrafast modulation of internal electric field [16], electron transport [17], laser output [18] and magnetic excitations [19-21]. In these works, the impact of picosecond strain pulses on a medium is governed by the deformation potential, piezoelectricity or magnetostriction. However, for realistic strain amplitudes (~$10^{-3}$), the low strength of these mechanisms means the practical applications of the technique are limited. The challenge in ultrafast strain engineering is either to develop methods to produce much higher strain on an ultrafast time scale, or to find mechanisms which provide a stronger strain-induced impact on electronic, structural and magnetic properties of nanostructures.

In the present Article we demonstrate experimentally an approach for ultrafast strain engineering, where picosecond strain pulses control ultrafast photo-induced phase transitions (PIPT) leading to a radical change of the media properties, e.g. dielectric susceptibility. The prerequisite for our work comes from the intensive studies of stationary strain- and stress-induced effects in nanoobjects fabricated from vanadium dioxide ($VO_2$) [22, 23]. Vanadium dioxide possesses an insulator-to-metal phase transition at close to room temperature ($T_c$=340 K for zero stress), and uniaxial stationary stress experiments have shown that it is a reliable material for strain nano-engineering [24, 25]. The excitation of $VO_2$ by intense femtosecond optical pulses induces ultrafast non-thermal PIPT [26] (for review see Ref. [27]), which has been also shown to be susceptible to stationary stress or strain [28-30]. These and other studies of PIPT point to $VO_2$ being a prospective material for experiments where ultrafast strain engineering could be realized by combining the impacts of picosecond strain pulses and pulsed optical excitation. Our experimental studies unambiguously demonstrate that a picosecond strain pulse of amplitude of <0.1% impacts ultrafast non-thermal PIPT. Strain pulse of the same amplitude has negligible effect on the phase transition dynamics at the timescale longer than ~10 ps after excitation, which is governed by temperature evolution.

The Article is organized as follows. In Sec. II we describe the main structural and optical properties of the epitaxial $VO_2$ nanohillocks grown on the $Al_2O_3$ substrate, and introduce the pump-probe technique designed for combined excitation of a medium by optical and strain pulses. In Sec. III we describe the PIPT driven in $VO_2$, by optical pulses and by combined action of the optical and strain pulses, as well as the effect the strain pulses alone have on the $VO_2$. It is followed by the extended analysis presented in Sec. IVA which shows that the strain pulses can indeed impede or enhance ultrafast PIPT in $VO_2$. In Sec. IVB we discuss a phenomenological model which qualitatively describes impact of picoseconds strain pulses on ultrafast PIPT, as well as on the nanosecond dynamics following PIPT. Conclusions and an outlook are presented in Sec. V.

## II. Experimental

### A. VO$_2$ nanohillocks on a Al$_2$O$_3$ substrate

The sample was a layer of epitaxial $VO_2$ nanohillocks grown on a 350 μm-thick *c*-plane sapphire, $Al_2O_3$, substrate by pulsed laser deposition [31]. Atomic force microscope (AFM) images [Fig. 1(a) and Fig. 1(b)] show that the hillocks have a height of 70±20 nm and lateral size of 200±55 nm. The $VO_2$ nanohillocks grown on *c*-cut $Al_2O_3$ are known to be single crystalline with the $[001]_M$-axis oriented in the plane of the sapphire substrate [21-33]. Figure 1(c) shows the temperature hysteresis of the optical reflectivity *R* at a photon energy of 1.2 eV, and reveals the phase transition occurring at $T_c$=340 K with coercivity of 20 K, which is typical for thin-film and nano-granular $VO_2$ samples [34, 35]. The changes of reflectivity from $R_i$ to $R_m$ at $T=T_c$ are due to the *changes of the refractive index* occurring when $VO_2$ undergoes the transition from insulating to metallic phase.

The sample was prepared for the experiments with picoseconds strain pulses, and a 140 nm thick Al film, serving as an opto-acoustic transducer [36], was sputtered on the back side of the sapphire.

### B. Combined optical-and-strain pump-probe setup

Figure 1(d) shows the pump-probe experimental scheme which allows combined excitation of a sample under study by femtosecond optical and picoseconds strain pump pulses. The laser source used is a 170 fs Yb:KGd(WO$_4$)$_2$ regenerative amplifier with a central photon energy of 1.2 eV and a repetition rate of 5 kHz. Each pulse from the source is split into three pulses. The first one, shown by red in Fig. 1(d), is the *optical pump pulse* with a fluence $W$ used to excite PIPT. The optical pump pulse is incident on $VO_2$ nanohillocks and is focused to a spot of a size of 55×100 μm². The *second pulse*, shown by blue, is used to generate the strain pulses [36]. It is incident onto an Al transducer with the fluence of ~60 mJ/cm² and the spot size of 110×110 μm². The third one, shown by dashed black line in Fig. 1(d) is the *probe pulse* controlled by a scanning delay line and used for monitoring the temporal evolution of the reflectivity $R(t)$ from the surface with $VO_2$ nanohillocks. For more details see Sec. III of the Supplemental Material [37].

We shall define the temporal reflectivity signals measured without and with strain pulses as $R_0(t)$ and $R_\varepsilon(t)$ respectively. The reflectivity $R_0(t)$ can take values between $R_i$ and $R_m$ which are the stationary values of reflectivity when all nanohillocks are in the insulating or metallic phases respectively. For the used photon energy excitation 1.2 eV (wavelength 1.03 μm) the maximum relative change of time dependent reflectivity in our sample is $(R_m-R_i)/R_i$ ~0.1.

**III. Experimental results**

**A. Ultrafast PIPT in $VO_2$**

Figure 1(e) demonstrates the pump-probe temporal reflectivity signal $R_0(\Delta t)$ in the absence of the strain pulse for three optical pump fluences, $W$ and $\Delta t=t-t_0$ ($t_0$ is the time when optical pump pulse is applied). The inset in Fig. 1(e) shows the dependence of reflectivity signal $R_0$ on $W$ at $\Delta t=1$ ps after the optical pulse impact on the $VO_2$. In agreement with earlier works [38-42] we see that the PIPT starts to take place above the threshold $W_T=6$ mJ/cm², and saturation is observed at $W>W_S=20$ mJ/cm² indicating that all $VO_2$ nanohillocks within the probe spot have undergone the PIPT under such pump fluence [43]. The wide range of $W$ between threshold $W_T$ and saturation $W_S$ results from a large dispersion of thresholds in the ensemble of nanohillocks with different sizes and other inhomogeneities [41]. The fraction of $VO_2$ which changes from insulating to metallic phase may be estimated from the fluence dependence of $R_0$ presented in Fig. 1(e). For instance, this fraction is about 50% for the excitation density $W=12.5$ mJ/cm².

**B. Generation of strain pulses and elasto-optical response of $VO_2$**

Following optical excitation of the Al transducer by the second pump (shown by blue in Fig.1(d)), strain pulses are injected into the sapphire substrate and propagate through it with the sound velocity (11 km/s) transforming into *N*-shape pulses due to the nonlinear elastic properties of sapphire [44], and reach $VO_2$ in 32 ns. An example of the simulated temporal strain profile, $\varepsilon_0(t)$, with strain amplitude ~$10^{-3}$ in sapphire in the vicinity of the $VO_2$ layer is shown in Fig. 1(f).

Figure 1(g) shows the simulated and measured evolutions of strain-induced reflectivity changes $\Delta r_0(t)$ in VO$_2$ in the absence of the optical pump ($W=0$). Further we designate the strain induced signal measured in the absence or presence of the pump beam exciting PIPT as $\Delta r_0(t)$ and $\Delta r_W(t)$ respectively. The temporal evolution of $\Delta r_0(t)$ is governed *only by the photo-elastic effect* in VO$_2$ and is proportional to the product of mean strain $\bar{\varepsilon}(t)$ in the VO$_2$ nanohillocks and the photoelastic constant $p$ ($p_i$ or $p_m$ in the insulating and metallic phases respectively). The signal $\Delta r_0(t)$ exhibits oscillatory behavior, and the temporal intervals where signal is positive and negative correspond to out-of-plane compression and tension respectively. The details of the strain and reflectivity simulations may be found in the Sec. I of Supplemental Material [37].

**C. Ultrafast PIPT in VO$_2$ under combined excitation by optical and strain pulses**

To examine the effect of a strain pulse on PIPT, we studied the reflectivity changes $\Delta r_W(t)$ of the VO$_2$ nanohillocks under simultaneous impact of both strain and optical pump pulses. The diagram in Fig. 1(h) shows the sequence of incident optical pulses and strain pulse on the sample. The delay $t_0$ is the time interval between the moments when the front edge of the strain pulse entered VO$_2$ and the optical pump pulse triggers PIPT. The value of $t_0$ is set to a specific value during the experiments. By changing the delay $t_0$ we induce PIPT during the strain pulse present in VO$_2$ (i.e. $t_0>0$) or before the strain pulse reaches the interface between sapphire and nanohillocks (i.e. $t_0<0$). Since the duration of the strain pulse when it reaches VO$_2$ is of ~100 ps (Fig. 1(e)), we can precisely adjust the temporal delay $t_0$ of the 170 fs optical pump in such a way that the latter excites the VO$_2$ nanohillocks during the action of out-of-plane compressive or tensile strain $\bar{\varepsilon}(t)$. The reflectivity changes are probed at a variable time $t$ which is counted from the moment the front edge of the strain pulse enters the VO$_2$ ($t = 0$).

The detection in our experiment is realized in a way that *only the strain-induced changes of the reflectivity are monitored*, either with or without the impact of the optical pump inducing PIPT, i.e. $\Delta r_0(t)$ or $\Delta r_W(t)$ respectively (for details see [37]). Then in the case of simultaneous excitation of VO$_2$ nanohillocks by optical pump and strain pulse the strain-induced probe signal $\Delta r_W(t)$ may be written as:

$$\Delta r_W(t) = p(t)\bar{\varepsilon}(t) + [R_\varepsilon(t, t_0) - R_0(t - t_0)] \quad (1)$$

There the first term describes the photo-elastic response proportional to the strain $\bar{\varepsilon}(t)$ in VO$_2$ and $p(t)$ is a corresponding photo-elastic constant. Both $\bar{\varepsilon}(t)$ and $p(t)$ depend on the phase, insulating or metallic, of VO$_2$, and thus for $W\neq0$ the elasto-optical constant is a time dependent function and depends on $W$ and $t_0$. Only in the case $W=0$ we get $\Delta r_W(t)=\Delta r_0(t)=p_i\bar{\varepsilon}(t)$, where $p_i$ is a photoelastic constant in the insulator phase. The second term in Eq. (1) which is the difference of the reflectivities with and without strain pulse [$R_\varepsilon(t,t_0)$ and $R_0(t-t_0)$ respectively] corresponds to the changes in the reflectivity governed by the changes *in refractive index due to PIPT* [for $R_0(t-t_0)$ see Fig. 1(e)]. $R_\varepsilon(t,t_0)$ and $R_0(t-t_0)$ have values between $R_i$ and $R_m$ and provide information on the VO$_2$ fraction transformed into metallic phase. The main goal of the experiments is to find the difference $\Delta R_\varepsilon(t,t_0)=R_\varepsilon(t,t_0)-R_0(t-t_0)$ associated with the strain-induced changes of a fraction which has experienced PIPT.

We obtain $\Delta R_\varepsilon(t,t_0)$ by subtracting the photoelastic contribution $p(t)\bar\varepsilon(t)$ in Eq. (1), from the measured $\Delta r_W(t)$. For this we start with the case when optical excitation $W$ exceeds the saturation level, $W_S$, and all VO$_2$ hillocks undergo PIPT to the metallic phase. The effect of the strain pulse on the PIPT in this case should be negligible, which means that $\Delta R_\varepsilon(t,t_0)=0$, and all changes in $\Delta r_W(t)$ are due only to the photoelastic effect. The results are presented in Fig. 2a. The main, red curve, represents the measured signal when the optical pulse excites VO$_2$ simultaneously with the strain pulse at the delay $t_0$=60 ps corresponding to the tensile part of the out-of-plane component of the strain pulse. A sudden change takes place in $\Delta r_W(t)$ at $t=t_0$ [for highly resolved temporal evolution see the inset in Fig. 2(a)]. The black curve corresponds to a signal $\Delta r_W(t) = \Delta r_0(t)=p_i\bar\varepsilon(t)$ at $W$=0, when all nanohillocks are in the insulating phase (see also Fig. 1g). The blue curve is the signal $\Delta\tilde r(t) = \Delta r_W(t)|_{t_0<0}$ obtained when the optical pulse hits the VO$_2$ before the arrival of the strain pulse. We have shown that for $W>W_S$, $\Delta\tilde r(t)$ is equal to $\Delta r_m(t)=p_m\bar\varepsilon(t)$ ($p_m$ is photoelastic constant in metallic phase) which is a reflectivity change when all VO$_2$ nanohillocks are in metallic phase (see Sec. II in Supplemental Material [37]). It is seen in Fig. 2(a) that at $t=t_0$=60 ps [see red curve in Fig. 2(a)], $\Delta r_W(t)$ switches abruptly from the photo-elastic response in the insulator phase [$\Delta r_W(t) =\Delta r_0(t)$ at $t<t_0$] to the response in the metallic phase [$\Delta r_W(t) =\Delta r_m(t)$ at $t>t_0$]. Two important conclusions can be drawn from this result. First, transient signals $\Delta r_W(t)=\Delta r_0(t)$ at all times $t<t_0$, confirming that VO$_2$ nanohillocks are in the insulator phase before the impact of optical pump. Second, transient signals $\Delta r_W(t) = \Delta\tilde r(t)\equiv\Delta r_m(t)$ at $t>t_0$. The transient around $t=t_0$ shown in the inset of Fig. 2(a) lasts less than 1 ps. Since $\Delta R_\varepsilon(t,t_0)=0$ for $W>W_S$ the effect of abrupt change of $\Delta r_W(t)$ at $t=t_0$ can be ascribed with confidence to the changes of $p$ upon the transition from an insulating to a metallic phase.

The case shown in Fig. 2(a) gives us a recipe for extracting $\Delta R_\varepsilon(t,t_0)$ for any $W$ which is the main aim of the experiment. This is done by comparing triads of signals $\Delta r_W(t)$, $\Delta r_0(t)$ and $\Delta\tilde r(t)$ measured for the same $W_T<W<W_S$ (for details see Sec. II in Supplemental Material [37]):

$$\Delta R_\varepsilon(t,t_0) = \begin{cases} \Delta r_W(t) - \Delta r_0(t) = 0, & t < t_0 \\ \Delta r_W(t) - \Delta\tilde r(t), & t > t_0 \end{cases} \qquad (2)$$

Now we turn to the most important part of the experiment where we measure $\Delta r_W(t)$ for intermediate optical fluences $W_T<W<W_S$ when a certain fraction of VO$_2$ nanohillocks undergoes PIPT. Figures 2(b) and 2(c) show corresponding triads: signals $\Delta r_W(t)$ with simultaneous excitation of PIPT and strain pulse (red curves); signals $\Delta r_0(t)$ for $W=0$ (black curves); and signals $\Delta \tilde{r}(t)$ measured when $t_0<0$ (blue curves). Again, the subpicosecond changes in transient reflectivity $\Delta r_W(t)$ take place at $t=t_0$ [see insets in Fig. 2(b)]. However, in strong contrast to data obtained at $W>W_S$ [Fig. 2(a)], at $t>t_0$ the transient reflectivity $\Delta r_W(t)$ clearly differs from $\Delta \tilde{r}(t)$, and thus $\Delta R_\varepsilon(t,t_0) \neq 0$ according to Eq. (2) after the pump pulse (i.e. at $t>t_0$). The signals $\Delta r_W(t)$ at $t>t_0$ are characterized not only by the reduced amplitude of oscillations, but these oscillations are superimposed on a different baseline. The latter effect is most evident at a nanosecond timescale [Fig. 2(c)], when the monotonously decaying behavior of $\Delta r_W(t)$ is clearly seen. At long time delays $t>300$ ps, when $\bar{\varepsilon}(t)=0$ in the VO$_2$ nanohillocks, the photo-elastic contribution in $\Delta r_W(t)$ [Eq. (1)] vanishes, leaving only the nonzero contribution $\Delta R(t,t_0)$ which is related to strain-induced changes of fraction of VO$_2$ which undergoes PIPT.

It is important that this slow decaying transient reflectivity $\Delta r_W(t)$ and, consequently, the nonzero $\Delta R_\varepsilon(t,t_0)$, is observed only for optical pump densities $W$ between the PIPT threshold $W_T$ and saturation values $W_S$. Furthermore, at elevated temperature $T=360$ K when all VO$_2$ is in metallic phase initially, no abrupt changes in $\Delta r_W(t)$ are detected at any $W$ and $t_0$ (see Sec. IV in Supplemental Material [37]).

## IV. Discussion

### A. Impact of the strain on ultrafast and nanosecond PIPT

The *main experimental result* of the present work is the observation of strain-induced sub-picosecond changes of the optical reflectivity associated with PIPT. These changes, defined in our work as $\Delta R_\varepsilon(t,t_0)$ are beyond the photoelastic effect, whose contribution may be subtracted from the measured signal $\Delta r_W(t)$ using the procedure described above. We attribute $\Delta R_\varepsilon(t,t_0)$ to the strain induced changes of VO$_2$ nanohillocks undergoing insulator-metal phase transition during PIPT. The main argument in favor of this statement is the observation of a long nanosecond decay of $\Delta R_\varepsilon(t,t_0)$ when the strain pulse in VO$_2$ is gone and there is no contribution from photo-elastic effect. The analysis of the transients $\Delta R_\varepsilon(t,t_0)$ obtained for different delays $t_0$ and different optical pump fluencies $W$ leads us to the following conclusions:

*(i)* Strain-induced decrease (increase) of the proportion of VO$_2$ nanohillocks undergoing PIPT takes place when out-of-plane compression (tension) takes place. In the experiments with strain pulses it is possible to control the sign and value of strain by choosing precisely the moment $t_0$ of optical impact inducing PIPT;

*(ii)* Only at the moment of the optical pulse impact the magnitude and sign of strain $\bar{\varepsilon}$ ($t_0$) define the strength of the strain-induced effect on PIPT and related quantitative difference $\Delta R_\varepsilon(t, t_0)$.

The conclusion *(ii)* means that the role of strain in PIPT is important only during ultrafast transients which include complex electron and lattice transformations and the presence of intermediate phases with sub-picosecond lifetimes (for review see Ref. [27]). Although our experiments do not allow us to distinguish whether strain affects mostly the electron or phonon systems, it is clear that strain does not have any effect on the state present on the longer timescales when recovery to the insulator phase is accompanied by thermal processes and sometimes metastable states with nanosecond transient time [42, 45]. It is important to stress that it has not been possible to make conclusion *(ii)* based on the experiments with stationary strain [30].

To support further the statements *(i)* and *(ii)* we plot in Fig. 3(c) transients $\Delta R_\varepsilon(t,t_0)$ vs. the reduced time $\Delta t=t-t_0$ at $W=9$ mJ/cm$^2$ and at two values of $t_0 = 60$ ps and 95 ps, corresponding to a maximum of out-of-plane compression and tension respectively. It is seen that there are two contributions to the decay of $\Delta R_\varepsilon(\Delta t)$ for both $t_0$ values: fast and slow, with respective amplitudes $A_F$ and $A_s$. The slow decay with time constant $\gtrsim 1$ ns is the relaxation of the material to the quasi-equilibrium state following PIPT, which lasts for microseconds, and most likely is governed by the local temperature equilibration in the excited spot [46]. Also the slow decay is not observed when PIPT is induced at $t_0$ close to the moment when $\Delta r_0(t)$ changes sign (see Sec. II in Supplemental Material [37]). These observations are consistent with the fact that the strain pulse does not change the temperature of the VO$_2$ and thus does not affect the processes underlying relaxation of the material to a quasi-equilibrium state at which $\Delta R_\varepsilon=0$.

Figures 3(b) and 3(c) show the optical pump fluence $W$ dependencies of $A_S$ and $A_F$ obtained as shown in Fig. 3(a). The results for $A_S$ confirm our main conclusions *(i)* and *(ii)*. Indeed $A_S \approx 0$ when $W<W_T$, and is also zero when $W$ exceeds the saturation level. From comparison of the measured relative changes of extracted $\Delta R_\varepsilon/R_i \sim 10^{-4}$ ($R_i$ – stationary reflectivity in insulator phase) and signal $\Delta R_0/R_i \sim 10^{-2}$ measured in the absence of the strain pulse we estimate the maximum additional fraction of VO$_2$ nanohillocks under input strain to be ~1% from the nanohillocks which undergo PIPT at $W=9$ mJ/cm$^2$. This estimate gives correct only to the order of magnitude because it is made under assumption of linear proportionality between the studiedlayer effective dielectric permittivity and fraction of material of nanohillocks in the metallic state. One can also expect the shift of the excitation threshold for PIPT under the strain pulse excitation. However, it is known from the experiments with stationary stress [30] that the 5 GPa stress is required to decrease the threshold $W_T$ to 0. In our experiments maximum stress in picosecond strain pulse is 0.1 GPa and then we may expect the threshold shift by about 1%, which agrees well with the maximum observed strain-induced change of PIPT. Such shift of $W_T$ cannot be clearly detected due to the fact that the onset of PIPT at $W_T$ is smeared due to inhomogeneity of nanohillocks' sizes, internal stresses etc.

The contribution $A_F$ of the fast decaying component in $\Delta R_\varepsilon(\Delta t, t_0)$ is negligible at $W<6$ mJ/cm$^2$, but it increases rapidly at higher $W$ and vanishes only at $W>26$ mJ/cm$^2$ which essentially exceeds $W_S$. Nonzero contribution from the fast decaying $\Delta R_\varepsilon(\Delta t, t_0)$ is observed also when PIPT is excited at $t_0$ close to the moment when $\Delta r_0(t)$ changes sign, and no slow contribution, attributed to the change of VO$_2$ fraction undergoing PIPT, is present (see Sec. II in Supplemental Material [37]). The origin of such behavior most likely is partly due to the difference of the elastic parameters of VO$_2$ in the insulator and metallic phases resulting in the dependence of $\bar{\varepsilon}(t)$ on the VO$_2$ phase [47]. It can also be related to the complex kinetics of the phase transition in the time interval $t=1 - 100$ ps [42, 45], for instance to the strain-induced changes of photoexcited carriers density reported in [28].

**B. Mechanism for strain-induced changes of ultrafast PIPT.**

The results described above unambiguously suggest that strain corresponding to out-of-plane compression (tension) applied at the moment of the optical pump impact, impedes (supports) the sub-picosecond insulator-to-metal transition and switching of the crystalline lattice to a new symmetry state (for VO$_2$ from monoclinic to rutile lattice cell) during PIPT. It is important that the impact of strain on the fraction of the VO$_2$ undergoing PIPT is observed only during a time less than 1 ps after the femtosecond optical pump pulse. Strain of the same amplitude acting on the VO$_2$ nanohillocks before the laser pulse impact does not have any effect on PIPT. Neither do strain applied at $t>1$ ps after optical pulse impact have any effect despite the system being in the metastable state [43].

To explain these findings on a qualitative level, we employ the phenomenological model of the first-order structural PIPT in a single domain of VO$_2$ [42, 48, 49] and extend it to the case of combined laser and strain pulse impacts. In this model, the thermodynamic potential $\Phi$ is introduced as a Landau expansion of the free energy [50] for the order parameter $\eta$:

$$\Phi(\eta) = \frac{\alpha(W,\varepsilon)}{2}\eta^2 + \frac{\beta}{4}\eta^4 + \frac{\gamma}{6}\eta^6, \qquad (3)$$

where the parameter $\alpha(W,\varepsilon) > 0$ is dependent on the exciting optical pump fluence and the applied strain, and $\beta < 0$ and $\gamma > 0$ are constants independent of $W$ and $\varepsilon$, values of which were found experimentally in [42]. Here, the order parameter $\eta$ is the generalized lattice distortion associated with lattice transformation from monoclinic to rutile phase, and $\Phi(\eta)$ is the energy of the system of two V atoms.

The present model considers the single domain nanoparticle and serves as a valid approximation for an individual nanohillock, which, most likely, either transits to a metallic phase or not as a whole [51]. However, if the size of a nanohillock allows coexistence of two phases within it, then Eq. (3) should be expanded with a term accounting for an energy penalty resulting from the emergence of the interphase boundary.

The thermodynamic potential $\Phi(\eta)$ for VO$_2$ in the equilibrium insulating monoclinic phase is shown in Fig. 4 by a solid black line. The value of $\eta_C$ corresponds to the equilibrium position of atoms in the insulating monoclinic phase and equals to the root mean square displacement of all atoms of VO$_2$ during the transition. Excitation by a femtosecond laser pulse drives the system into nonequilibrium state which is characterized by the presence of two minima in $\Phi(\eta)$. These minima correspond to the laser-induced rutile phase at $\eta = 0$, and the metastable monoclinic phase at $0 < \eta < \eta_C$ [40]. The black dashed line in Fig 4(a) shows an example of $\Phi(\eta)$ for photoexcited VO$_2$ in the case of moderate optical pump fluence $W_T < W < W_S$. Excitation of the nonequilibrium state triggers the structural phase transition which proceeds in two steps [42, 43]: At the initial stage, the laser pulse fluence $W_T < W < W_S$ is sufficient for an over-barrier excitation, and yields partial transitions to the both minima corresponding to the rutile and metastable monoclinic phases. The fraction of VO$_2$ in the rutile phase after excitation is determined by the height of the barrier $\Delta G$ [see inset in Fig. 4(a)]. This initial stage for structural PIPT is governed by coherent optical phonons and develops at a time $\tau_1 < \sim 1$ ps [42, 43]. After that the system appears either in the rutile phase or in a potential well of the metastable monoclinic state at $0 < \eta < \eta_C$. The second stage includes slow ($\tau_2 > 1$ ns) thermally activated transitions over the barrier $\Delta G$ from the metastable monoclinic to rutile phase and a final cooling approaching the equilibrium monoclinic phase.

The applied strain changes the parameter $\alpha(W,\varepsilon)$ in the thermodynamic potential $\Phi(\eta)$ of the photoexcited system, given by the Eq. (3) [see red and blue lines in Fig. 4(a)] [52]. Since PIPT is complete at the timescale, which is significantly shorter than the characteristic time of strain modulation in our pulse, the system during PIPT is affected by the strain as if it was a quasi-stationary one, with a particular magnitude and sign. Therefore, the strain increases or decreases the over-barrier excitation energy [see vertical arrows in Fig. 4(a)], and thus alters the fraction of VO$_2$ in the rutile phase at the first (i.e. picosecond) stage of PIPT, as indicated by the symbols in Fig. 4(a). Such a model explains, on a qualitative level, the experimentally observed strain-induced effect on PIPT during $t\sim\tau_1$ following optical excitation.

The second (i.e. nanosecond) stage of PIPT should also be sensitive to the strain due to the strain modulation of the barrier height $\Delta G$ [see inset in Fig. 4(a)]. However, in contrast to the impact of strain at the first ultrafast stage of PIPT, the slow over-barrier transition can be efficiently modulated by strain only if the latter is applied during time $t>\sim\tau_2$. It is easy to show that the amplitude of the strain-induced modulation of the rutile phase at this stage is proportional to $(\omega\tau_2)^{-1}$, where $\omega$ is a characteristic radial frequency of the coherent acoustic phonon wavepacket in the strain pulse. In our experiments $\omega\sim 10^{11}$ rad·s$^{-1}$ and $(\omega\tau_2)^{-1}\leq 10^{-2}$, which means that modulation by strain is two orders of magnitude more efficient at the first ultrafast stage of PIPT than at the second nanosecond stage when over-barrier processes are required for structural phase transition. This accounts well for the absence of the strain-induced modulation on the PIPT at timescales longer that 1 ps after the optical excitation [Fig. 3(a)].

In the saturation regime, when the laser pulse fluence $W>W_S$, the minimum in $\Phi(\eta)$ corresponding to the metastable monoclinic phase vanishes, and the complete PIPT to the rutile phase occurs already at the first ultrafast stage of PIPT [Fig. 4(b)]. In this case, moderate strain applied to VO$_2$ is not sufficient to introduce the second minimum in $\Phi(\eta)$ at $0 < \eta < \eta_C$ and, thus, PIPT is insensitive to the impact of strain pulses. This is consistent with our experimental observations.

Finally, we note that the symmetry of the VO$_2$ nanohillocks used in the experiments requires in-plane strain components [22, 23] to influence the lattice switching. The injected strain components $\varepsilon_{zz}$ ($z$ is a direction perpendicular to the surface plane) are out of plane, but in-plane components are generated in the VO$_2$ hillocks if their diameter is not much larger than their height [53, 54]. This is indeed the case for the studied VO$_2$ nanohillocks, which have a diameter-to-height ratio of ~3 [Fig. 1(a)]. Thus, we argue that in our experiments, the tensile (compressive) in-plane strain reduces (increases) the fraction of VO$_2$ undergoing PIPT, which is in agreement with the static experiments [22, 23].

## V. Conclusions

We have shown that the impact of picosecond strain pulses with amplitude ~10$^{-3}$ decreases or increases, depending on the sign of strain, the fraction of VO$_2$ nanohillocks which undergo ultrafast PIPT from insulating to metallic phase. This impact occurs only at a sub-picosecond time range after optical pulse excitation. After the strain pulse the relaxation of the excess or deficient fraction of $VO_2$ in the metallic phase to the quasi-equilibrium takes place in a nanosecond time scale, which is faster than for full recovery of $VO_2$ from PIPT.

The observed ~1% change of strain-induced modulation of the $VO_2$ volume undergoing the phase transition may be significantly enhanced by increasing the picosecond strain amplitude from 0.1% up to state-of-the-art values of ~1.5% [55, 56]. Furthermore, in a single domain nanoobject, e.g. a single nanohillock, the excitation threshold for PIPT will be well defined and not spread over the wide range of optical intensities. In this case, the compressive and tensile components of the picosecond strain pulse should notably increase or decrease the threshold value for PIPT. In such a system of well-defined nanoelements the picosecond-strain-assisted enhancement or suppression of the ultrafast PIPT may lead to prospective applications in CMOS and photonic technologies [57-60]. Since dynamical strain may be localized down to nanometer scale [61], it can be used as a tool for selective control of single $VO_2$ nanoelements. One can envisage an all-optically controlled nanoarray of ultrafast electrical/optical switches, where the optical excitation drives the transition to the metallic state selectively in an element of the array which is subject to dynamical strain at the moment of excitation.

The demonstrated effect is not limited to the particular material and type of phase transitions studied here. Feasibility of the control of PIPT in $VO_2$ by picosecond strain pulses paves the way to ultrafast strain engineering in materials with magnetic phase transitions where femtosecond photo-induced changes of magnetic state have been revealed [62-65]. Ultrafast strain engineering may also facilitate yet-to-be-demonstrated laser-driven control of ferroelectricity in complex structures, i.e. heterostructures and patterned nanolayers, which include optically opaque and transparent materials possessing phase transitions. Picosecond strain pulses may be generated selectively in space, thus allowing control of strain-induced effects on both nanometer and picosecond scales in space and time respectively.

**Acknowledgements**

The collaboration between the Ioffe Institute and the University of Nottingham was supported by the RFBR grant No. 17-52-10015 and the Royal Society grant No. IEC\R2\170217. The experiments were performed under support of the RScF grant No. 16-12-10520 at the Ferroics Physics Laboratory. S.L. and F.F. were supported in part by the U. S. Army Research Laboratory and the U. S. Army Research Office under contract No. W911NF-15-1-0448. The collaboration between the Ioffe Institute and TU Dortmund was supported by DFG through ICRC TRR160 (project B6) and RFBR through the grant No. 15-52-12015 and the Volkswagen Foundation (grant No. 90418). We thank A. S. Salasyuk and A. E. Fedianin for help with the experiments, S. M. Suturin and R. M. Dubrovin for help with the AFM measurements, B. A.12

**Figure Captions**

Figure 1. (Color online) Sample properties and experimental schematic. (a) AFM image of the VO$_2$ nanohillocks and (b) its cross-section along the blue line in (a). (c) The temperature dependence of stationary reflectivity $R$ at the photon energy 1.2 eV obtained upon heating (red symbols) and cooling (blue symbols). (d) Experimental scheme. (e) Transient reflectivity $R_0(\Delta t)$ signals obtained for three values of optical fluence $W$ in the absence of the strain pulse. Inset shows the transient reflectivity at $t'$=1 ps as a function of $W$. (f) Calculated strain pulse temporal profile $\varepsilon(t)$ in nonlinear propagation regime in the sapphire substrate in the vicinity of the VO$_2$ nanohillocks; (g) Calculated (blue line) and measured (red line) reflectively evolution of strain-induced reflectivity signal $\Delta r_0(t)$ in the absence of the optical pump ($W$=0) corresponding to nonlinear input strain pulse shown in (f). (h) The diagram illustrating definitions of $t$=0, $t$=$t_0$, and $\Delta t$, and the instants for excitations of Al transducer and PIPT in VO$_2$.

Figure 2. (Color online) Temporal evolutions of the reflectivity changes in the VO$_2$ nanohillocks under the impact of the picosecond strain pulse measured in the time intervals 300 ps (a, b) and 1700 ps (c). Black and red curves are the signals measured without [$\Delta r_0(t)$] and with [$\Delta r_W(t)$] optical pump. Vertical arrows indicate the time $t_0$> 0 when the optical pump is applied. Blue lines correspond to $\Delta \tilde{r}(t)$ when the optical pulse excited the VO$_2$ nanohillocks before the arrival of the strain pulse, $t_0$=-30 ps. In (b) and (c) the optical pump densities $W$ are above the threshold, $W$>$W_T$, and below the saturation level, $W$<$W_S$, for PIPT; in (a) $W$>$W_S$. The insets in (a) and (b) show $\Delta r(t)$ vs. reduced time $\Delta t$=$t$-$t_0$ measured around $t_0$ with temporal resolution 200 fs. Also indicated are the mean strain amplitudes $\bar{\varepsilon}$ at the moment of the photoexitation $t_0$ (See for details Sec. I and Figure S5 in Supplementary Material [37]).

Figure 3. (Color online) Strain-induced effect in the ultrafast phase transition. (a) Illustration of the procedure [Eq. (2)] for extracting the contribution $\Delta R_\varepsilon(t, t_0)$ due to strain present during the photoexcitation (at $t_0$= 60 ps) from the total signal $\Delta r_w(t)$ measured at $t$=65 ps. The black and red curves are the signals measured without [$\Delta r_0(t)$] and with [$\Delta r_W(t)$] optical pump. Vertical arrows indicate the time $t_0$=60 ps when the optical pump is applied. Blue lines correspond to $\Delta r(t)$ when the optical pulse excited the VO$_2$ nanohillocks before the arrival of the strain pulse, $t_0$=-30 ps [see also Fig. 2(b)].

(b) Expanded view of frame (a) for $t_0$=60 ps (upper panel) and $t_0$=95 ps (lower panel) in the time interval of 300 ps. Shaded areas show $\Delta R_\varepsilon(t, t_0)$ and highlights that the sign of this strain-induced contribution is conserved over the whole temporal range $t>t_0$. (c) Temporal evolutions of the extracted contribution $\Delta R_\varepsilon(t)$ to the PIPT induced by out-of-plane tensile ($\bar{\varepsilon}(t_0$=60 ps$) \approx 1.3\times10^{-3}$, upper) and compressive (at $\bar{\varepsilon}(t_0$=95 ps$) \approx -2\times10^{-3}$, lower) strain. (d, e) Optical pump fluence dependences of the amplitudes of the fast $A_F$ (open circles) and slow $A_S$ (closed triangles) components of $\Delta R_\varepsilon(t)$ as obtained for out-of-plane tensile (d), and compressive (e) strain. Solid lines are the guides to the eye. Vertical dashed lines mark the PIPT threshold $W_T$ and saturation $W_S$ optical pump fluencies.

Figure 4. (Color online) Thermodynamic potentials $\Phi(\eta)$ [Eq. (3)] in the initial monoclinic phase (solid lines) and after the photoexcitation (dashed lines) by a femtosecond pulse of intermediate fluence (a), and in the saturation regime (b), calculated using the parameters determined in [42]. Black lines show the thermodynamic potential of the unstrained system. Red and blue lines are the potentials corresponding to the strain components which increase ($\varepsilon$+) or decrease ($\varepsilon$-) the free energy of the photoexcited system, respectively. Note that the distortion of $\Phi(\eta)$ due to strain is exaggerated for the sake of clarity. Vertical arrows show the photoexcitation process in the sample in the presence of strain. Symbols and their sizes indicate schematically a probability for the system to occur in the rutile ($\eta$=0) or metastable monoclinic ($0<\eta<\eta_C$) state after ~1 ps following the photoexcitation of the sample subjected to the strain-induced increase (red circles) or decrease (blue squares) of $\Phi(\eta)$ at $\eta=\eta_C$. Inset shows the strain-induced change of the potential barrier $\Delta G$ between the photoexcited rutile and metastable monoclinic states.

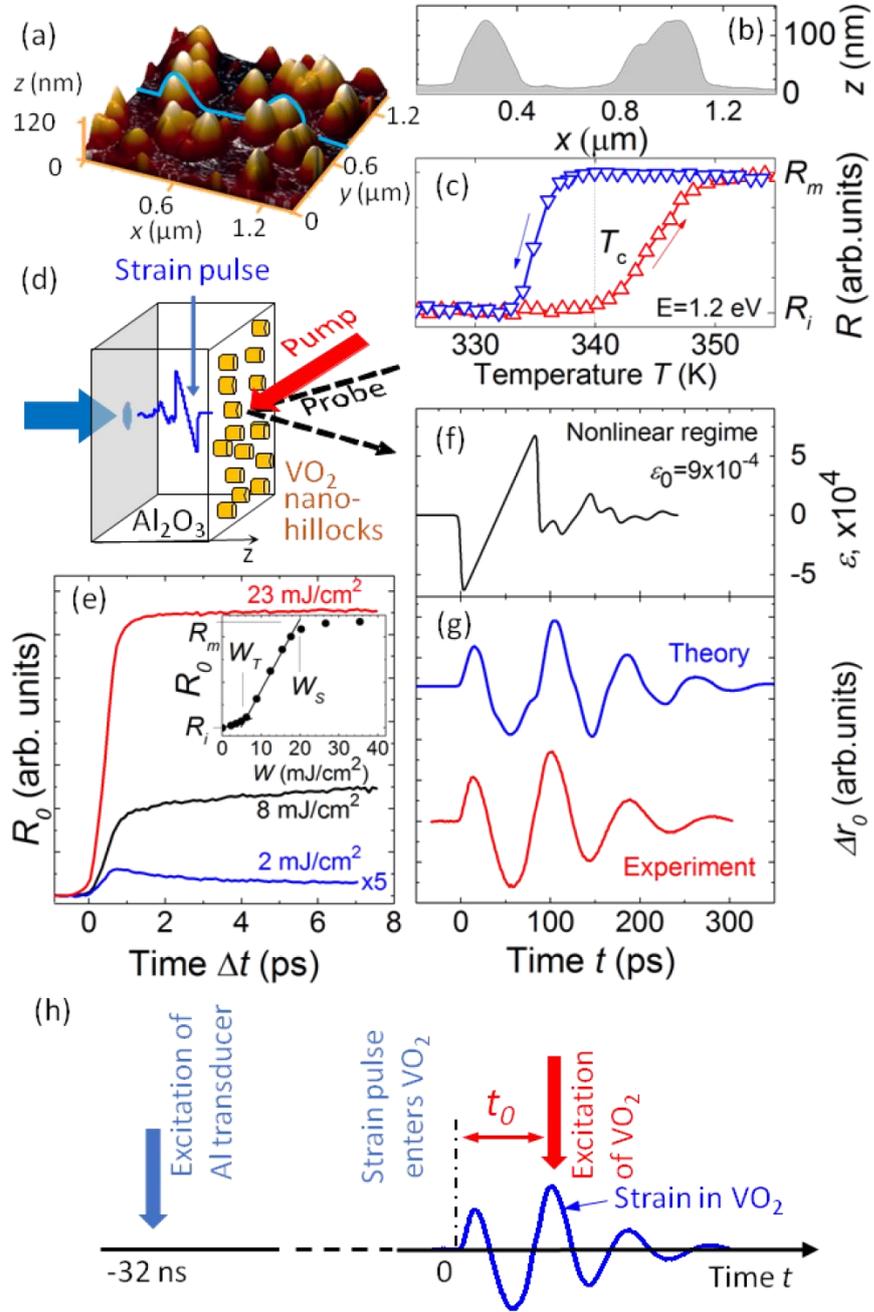

Fig. 1 (Color online) Sample properties and experimental schematic. (a) AFM image of the VO$_2$ nanohillocks and (b) its cross-section along the blue line in (a). (c) The temperature dependence of stationary reflectivity $R$ at the photon energy 1.2 eV obtained upon heating (red symbols) and cooling (blue symbols). (d) Experimental scheme. (e) Transient reflectivity $R_0(\Delta t)$ signals obtained for three values of optical fluence $W$ in the absence of the strain pulse. Inset shows the transient reflectivity at $t'=1$ ps as a function of $W$. (f) Calculated strain pulse temporal profile $\varepsilon(t)$ in nonlinear propagation regime in the sapphire substrate in the vicinity of the VO$_2$ nanohillocks; (g) Calculated (blue line) and measured (red line) reflectively evolution of strain-induced reflectivity signal $\Delta r_0(t)$ in the absence of the optical pump ($W=0$) corresponding to nonlinear input strain pulse shown in (f). (h) The diagram illustrating definitions of $t=0$, $t=t_0$, and $\Delta t$, and the instants for excitations of Al transducer and PIPT in VO$_2$.

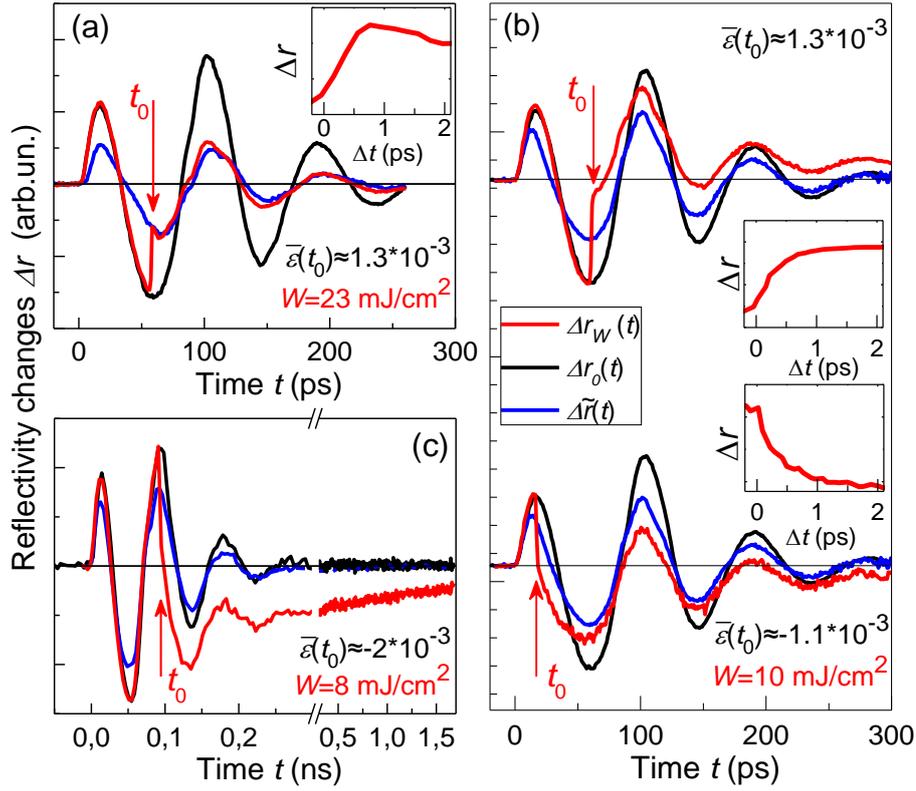

Fig. 2 (Color online) Temporal evolutions of the reflectivity changes in the VO$_2$ nanohillocks under the impact of the picosecond strain pulse measured in the time intervals 300 ps (a, b) and 1700 ps (c). Black and red curves are the signals measured without [$\Delta r_0(t)$] and with [$\Delta r_W(t)$] optical pump. Vertical arrows indicate the time $t_0 > 0$ when the optical pump is applied. Blue lines correspond to $\Delta \tilde{r}(t)$ when the optical pulse excited the VO$_2$ nanohillocks before the arrival of the strain pulse, $t_0 = -30$ ps. In (b) and (c) the optical pump densities $W$ are above the threshold, $W > W_T$, and below the saturation level, $W < W_S$, for PIPT; in (a) $W > W_S$. The insets in (a) and (b) show $\Delta r(t)$ vs. reduced time $\Delta t = t - t_0$ measured around $t_0$ with temporal resolution 200 fs. Also indicated are the mean strain amplitudes $\bar{\varepsilon}$ at the moment of the photoexitation $t_0$ (See for details Sec. I and Figure S5 in Supplementary Material [37]).

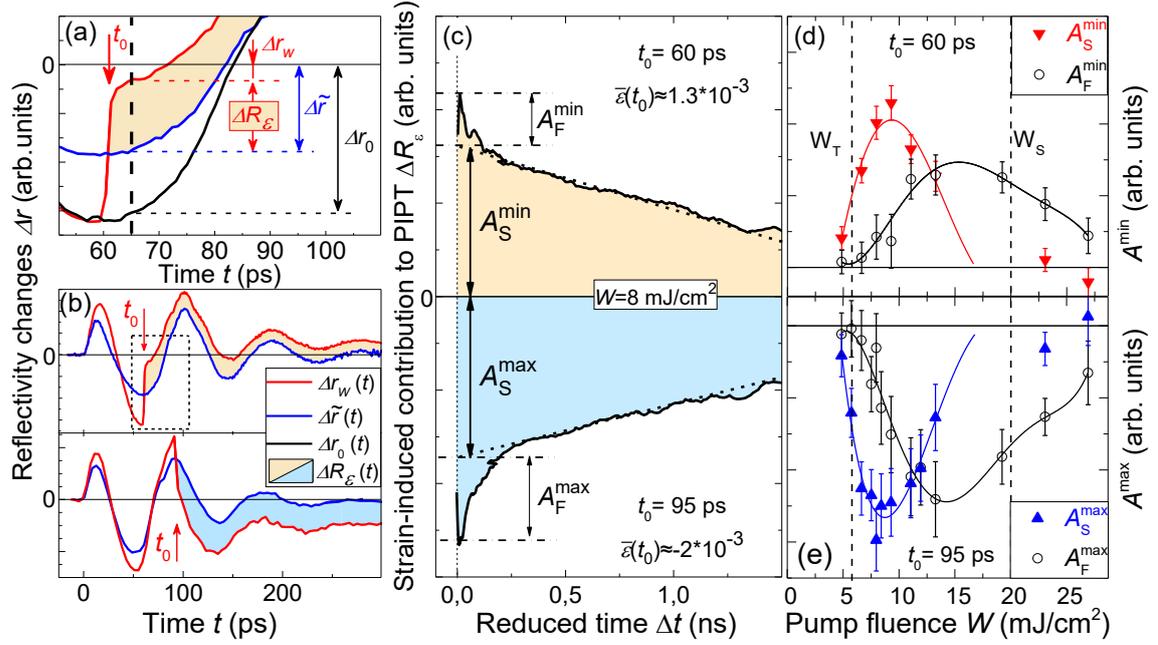

Fig. 3 (Color online) Strain-induced effect in the ultrafast phase transition. (a) Illustration of the procedure [Eq. (2)] for extracting the contribution $\Delta R_\varepsilon(t, t_0)$ due to strain present during the photoexcitation (at $t_0$= 60 ps) from the total signal $\Delta r_w(t)$ measured at $t$=65 ps. The black and red curves are the signals measured without [$\Delta r_0(t)$] and with [$\Delta r_W(t)$] optical pump. Vertical arrows indicate the time $t_0$=60 ps when the optical pump is applied. Blue lines correspond to $\Delta r(t)$ when the optical pulse excited the VO$_2$ nanohillocks before the arrival of the strain pulse, $t_0$=-30 ps [see also Fig. 2(b)]. (b) Expanded view of frame (a) for $t_0$=60 ps (upper panel) and $t_0$=95 ps (lower panel) in the time interval of 300 ps. Shaded areas show $\Delta R_\varepsilon(t, t_0)$ and highlights that the sign of this strain-induced contribution is conserved over the whole temporal range $t>t_0$. (c) Temporal evolutions of the extracted contribution $\Delta R_\varepsilon(t)$ to the PIPT induced by out-of-plane tensile ($\bar{\varepsilon}(t_0$=60 ps) ≈ 1.3×10$^{-3}$, upper) and compressive (at $\bar{\varepsilon}(t_0$=95 ps) ≈ -2×10$^{-3}$, lower) strain. (d, e) Optical pump fluence dependences of the amplitudes of the fast $A_F$ (open circles) and slow $A_S$ (closed triangles) components of $\Delta R_\varepsilon(t)$ as obtained for out-of-plane tensile (d), and compressive (e) strain. Solid lines are the guides to the eye. Vertical dashed lines mark the PIPT threshold $W_T$ and saturation $W_S$ optical pump fluencies.

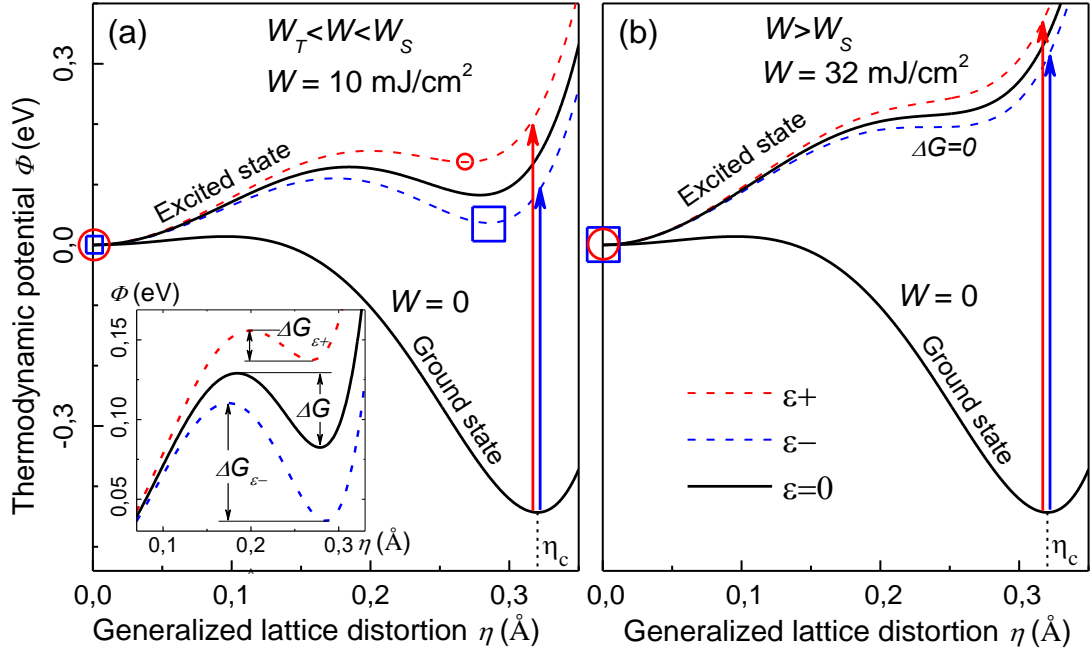

Fig. 4. (Color online) Thermodynamic potentials $\Phi(\eta)$ [Eq. (3)] in the initial monoclinic phase (solid lines) and after the photoexcitation (dashed lines) by a femtosecond pulse of intermediate fluence (a), and in the saturation regime (b), calculated using the parameters determined in [42]. Black lines show the thermodynamic potential of the unstrained system. Red and blue lines are the potentials corresponding to the strain components which increase ($\varepsilon+$) or decrease ($\varepsilon-$) the free energy of the photoexcited system, respectively. Note that the distortion of $\Phi(\eta)$ due to strain is exaggerated for the sake of clarity. Vertical arrows show the photoexcitation process in the sample in the presence of strain. Symbols and their sizes indicate schematically a probability for the system to occur in the rutile ($\eta=0$) or metastable monoclinic ($0<\eta<\eta_C$) state after ~1 ps following the photoexcitation of the sample subjected to the strain-induced increase (red circles) or decrease (blue squares) of $\Phi(\eta)$ at $\eta=\eta_C$. Inset shows the strain-induced change of the potential barrier $\Delta G$ between the photoexcited rutile and metastable monoclinic states.

# Ultrafast insulator-to-metal transition in $VO_2$ nanostructures assisted by picosecond strain pulses

Ia.A. Mogunov,[1,*] F. Fernández,[2] S. Lysenko,[2] A.J. Kent,[3] A.V. Scherbakov,[1,4]
A.M. Kalashnikova,[1] and A.V. Akimov[3]

[1] *Ioffe Institute, 194021 St. Petersburg, Russia*

[2] *Department of Physics, University of Puerto Rico, Mayaguez, 00681 Puerto Rico, USA*

[3] *School of Physics and Astronomy, University of Nottingham, Nottingham NG7 2RD, UK*

[4] *Experimentelle Physik 2, Technische Universität Dortmund, D-44227 Dortmund, Germany*

[*] *corresponding author: mogunov@mail.ioffe.ru*

## I. Strain pulses and photoelastic effect in $VO_2$ nanohillocks

### A. Strain pulse propagation in sapphire

The strain pulses in our experiments are generated due to the thermoelastic effect in the Al film deposited on the back side of the sapphire substrate: the optical excitation from the pulsed laser transmitted into the Al film is absorbed there. The temperature rise of the film generates stress and this stress launches the strain pulse which could be considered as a wave packet of coherent longitudinal acoustic phonons. The details of strain pulse generation in opaque films are described in detail elsewhere [C. Thomsen, H. T. Grahn, H. J. Maris, and J. Tauc, Surface generation and detection of phonons by picosecond light pulses, Phys. Rev. B **34**, 4129-4138 (1986)].

The part of strain pulse generated in the Al film is injected into sapphire (c-cut). Quite a large acoustic mismatch between Al and sapphire yields a reflection coefficient $R=0.44$ at the interface and leads to multiple reflections of the strain pulse in the Al film. This has allowed us to measure the echo pulse in the Al film (see Fig. S1) in the transient optical reflectivity and get precisely the thickness of the Al film $d=138$ nm.

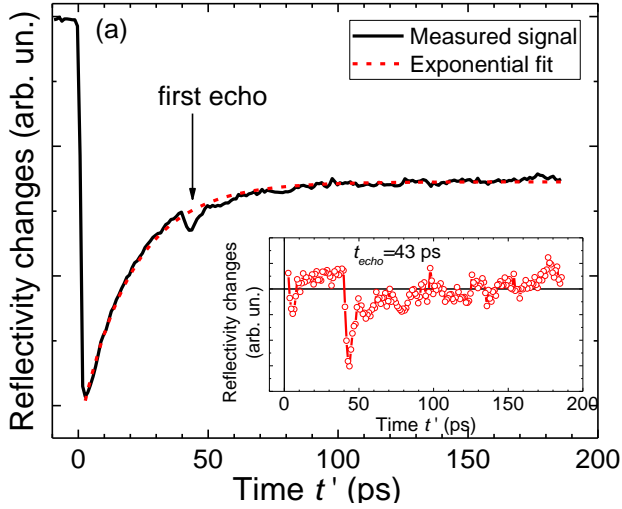

Figure S1. (a) Optical pump-probe signal (black solid line) measured from the Al film on the back of the sapphire substrate. The echo is indicated by the vertical arrow. The inset shows the difference between the signal and its fit to a single exponential decay function (red dashed line).

The temporal shape of the initial strain pulse can be modeled using the Gaussian shape for displacement $u(z-ct)$ in the Al film. Then, taking into account multiple reflections of the strain pulse in the film, we may simulate the temporal evolution of the strain pulse $\varepsilon = \frac{du}{dz}$ propagating through sapphire substrate. The temporal shape of $\varepsilon_0(t)$ injected into sapphire substrate is shown in Fig. S2.

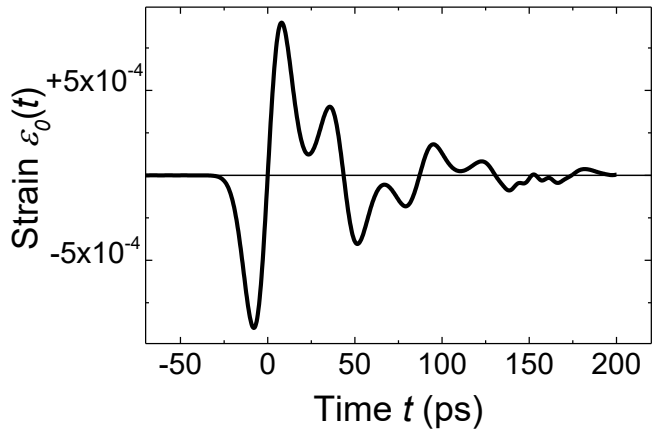

Figure S2: The calculated temporal shape of the strain pulse $\varepsilon_0(t)$ with amplitude $\varepsilon_0 = 9\times10^{-4}$ injected from the metal film into the sapphire.

The amplitude of injected strain pulse and the distance of 350 μm traveled in sapphire are large enough for the nonlinear acoustic propagation in the sapphire to have a pronounced effect [P.J.S. van Capel, E. Péronne, and J.I. Dijkhuis, Nonlinear ultrafast acoustics at the nanoscale, Ultrasonics **56**, 36 (2015)]. The result for the simulated strain pulse in sapphire with the amplitude $10^{-3}$ is shown in Fig.S3 (see also Fig. 1(f) in the main text).

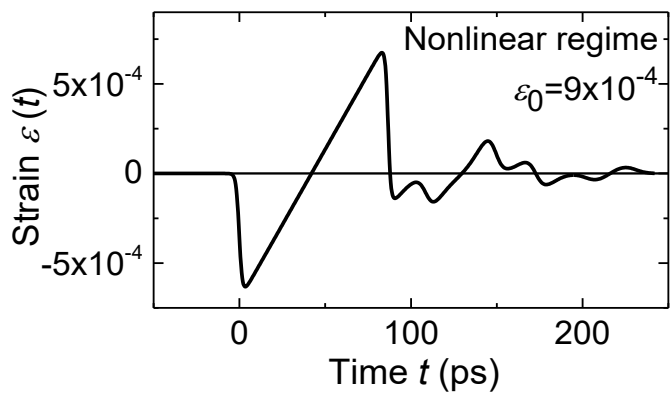

Fig.S3. Calculated strain pulse temporal profile in the nonlinear propagation regime in the sapphire substrate in vicinity of VO$_2$ nanohillocks for initial strain amplitude $\varepsilon_0 = 9\times10^{-4}$.

*B. Strain pulse and photoelastic effect in $VO_2$ nanohillocks*

The simulations of the temporal evolution of the photoelastic response $\Delta r_0(t)$ to the reflectivity signal have been done in two steps. The first step is the calculation of the strain evolution as a function of time and coordinate in $VO_2$, and the second is the calculation of reflectivity changes in the presence of a strain distribution calculated in the first step.

The strain pulse in sapphire has only a longitudinal component of the strain tensor $\varepsilon_{zz}(t,z)$. In the general case such a strain pulse hitting $VO_2$ nanohillocks excites there all strain components $\varepsilon_{ij}(t,z)$, where $i,j = x, y, z$. The strain evolution in this case depends on the particular shape of the nanohillock and the analysis of the reflectivity signal becomes a complicated task with unknown elastic constant and photoelastic tensors.

We find that for qualitative analysis of the photoelastic response in the present case it is possible to simplify the analysis and compare the experimental results with calculations of reflectivity induced by the strain pulse in a plane $VO_2$ film with the thickness $a\sim70$ nm which corresponds to the average height of nanohillocks, and an effective sound velocity $v_{eff}$. The validity of such a simplification is verified by a good qualitative agreement of the calculated and measured reflectivity signals $\Delta r_0(t)$ (see Fig. 1 (g) in the main text).

For the simulation we used MATLAB codes designed by Thomas Czerniuk (Dortmund University) for calculations of the photoelastic effect in multilayer structures [https://eldorado.tu-dortmund.de/bitstream/2003/36067/1/Dissertation_Czerniuk_pdfA.pdf]. The codes are based on the transfer matrix formalism for phonons and photons in the first and second stages of the calculations for strain pulse evolution and photoelastic effect respectively. The latter includes the displacement of interfaces and changes of the refractive index $n$ in the $VO_2$ film. We used the following parameters:

Refractive index of $n(VO_2)$ at $\lambda=1.03\mu$m is $n=3+0.5i$ [H. W. Verleur, A. S. Barker, Jr., C. N. Berglund, Optical Properties of VO2 between 0.25 and 5 eV, Phys. Rev., **172**, 788 ((1968))]. Refractive index of sapphire is 1.8; density of $VO_2$ is 4340 kg/m$^3$.

Figure S4 shows the normalized calculated shapes of $\Delta r_0(t)$ for photoelastic constant $p=-2$ and four values of effective sound velocities $v_{eff}$. We get the best agreement when $v_{eff}=3500$ m/s in the insulating phase of $VO_2$. It is important to mention that the value of $v_{eff}$ is not equal to a velocity of any particular phonon mode measured and calculated in earlier works [E. Abreu, S. N. Gilbert Corder, S. J. Yun, S. Wang, J. G. Ramírez, K. West, J. Zang, S. Kittiwatanakul, I. K. Schuller, J. Lu, *et al*, Ultrafast electron-lattice coupling dynamics in VO$_2$ and V$_2$O$_3$ thin films, Phys. Rev. B **96**, 094309 (2017)]. The $v_{eff}$ is an average value governed by longitudinal and transverse phonon modes and depends on the ratio between the height and diameter of the hillocks.

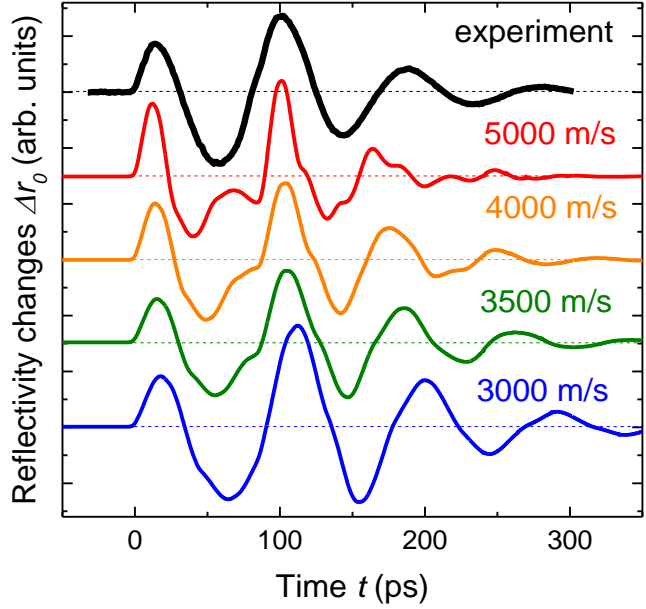

Figure S4. Calculated (colored lines) and measured (black line, see also Fig.1g of the main text) evolution of strain-induced reflectivity signal $\Delta r_0(t)$ in the absence of the optical pump. The calculated curves were obtained for different values of velocity $v_{eff}$ in $VO_2$, and the values of $v_{eff}$ are shown above each curve.

The calculated photoelastic response with optimized parameters (Figure S4 and Figure 1 (g) of the main text) allowed us to estimate the mean strain amplitude in $VO_2$ nanohillocks at the moments $t_0$ when PIPT was excited. Figure S5 shows strain profiles $\varepsilon(z,t=t_0)$ within the $VO_2$ as derived from this calculation. The spatial coordinate range corresponds to the 70 nm film thickness chosen as an average height of nanohillocks. Since light penetration depth in $VO_2$ at $\lambda=1030$ nm exceeds the thickness of the simulated film, the mean strain amplitude which contributes to the optical response was calculated as an average value within the film. The corresponding values of mean strain give: $\bar{\varepsilon}(t_0)=1.3\times10^{-3}$ for the first minimum of $\Delta r_0(t)$ (tensile out-of-plane strain) and $\bar{\varepsilon}(t_0)=-2\times10^{-3}$ for the second maximum of $\Delta r_0(t)$ (compressive out-of-plane strain).

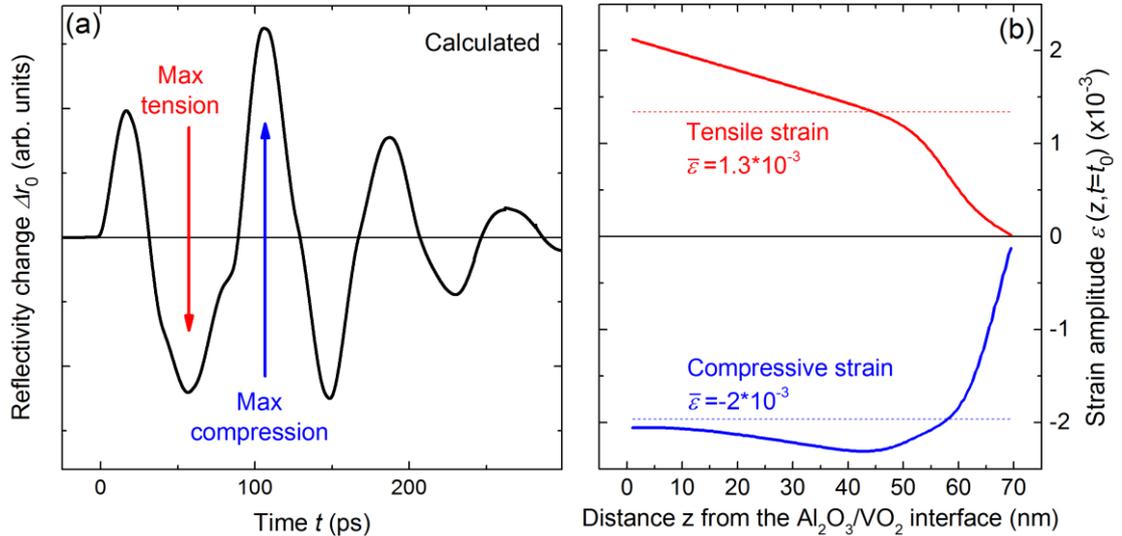

Figure S5: (a) Calculated reflectivity change $\Delta r_0(t)$. The instants in time of the maximum tension and compression are marked with arrows. During the experiments these instants were chosen as $t_0$ to drive PIPT during the maximal out-of-plane tension or compression in $VO_2$ nanohillocks. (b) Calculated spatial profiles of $\varepsilon(z,t=t_0)$, $z=0$ is sapphire/$VO_2$ interface and $z=70$ nm is $VO_2$/air boundary. For the red line $t_0$ corresponds to the most out-of-plane tension, and for the blue line $t_0$ stands as a moment of maximal out-of-plane compression. The dashed lines show average strain $\bar{\varepsilon}(t_0)$.

## II. Subtraction of photoelastic contribution and extracting $\Delta R_\varepsilon(t,t_0)$

Eq.(2) in the main text includes the signal $\Delta \tilde{r}(t)$ measured when the optical pulse hits the $VO_2$ before the arrival of the strain pulse, i.e. when $t_0<0$. The signal $\Delta \tilde{r}(t)$ represents purely photoelastic contribution to $\Delta r_\varepsilon(t)$ at $t>t_0$. Figure S6 shows the reflectivity changes $\Delta \tilde{r}(t)$ in the $VO_2$ nanohillocks at $t_0=-40$ ps (blue curve) and $t_0=-5$ ps (green curve). The excitation density $W=10$ mJ/cm$^2$ of the optical pump used is above the threshold and below the saturation level for photoinduced phase transition in $VO_2$, $W_T<W<W_S$. The black curve shows the signal $\Delta R_0(t)$ measured in the absence of optical pump.

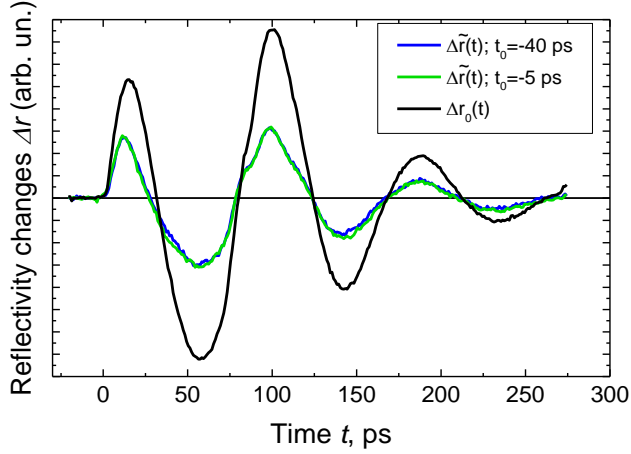

Figure S6. Temporal evolutions of reflectivity changes $\Delta \tilde{r}(t)$ in $VO_2$ hillocks under the impact of picosecond strain pulse after optical pump excitation, i.e. with negative delay between them $t_0<0$: $t_0=-40$ ps (blue curve) and $t_0=-5$ ps (green curve). The excitation density $W=10$ mJ/cm$^2$ of the optical pump is above the threshold and below the saturation level for photo-induced phase transition in $VO_2$, $W_T<W<W_S$.

It is seen that the effect of the optical pump is a decrease of the oscillation amplitude in the reflectivity response and a slight change of the temporal shape. Comparing the signals $\Delta \tilde{r}(t)$ in Fig.S6 with the signals $\Delta r(t)$ in Fig.2 of the main text we see that $\Delta \tilde{r}(t)$ does not show any shift of the base line and slow decay which is clearly detected when the optical pump hits $VO_2$ at $t_0>0$ corresponding to nonzero strain $\bar{\varepsilon}(t_0)\neq 0$. This means that the strain pulse under conditions $t_0<0$ does not change the fraction of $VO_2$ in metallic phase and correspondingly $\Delta R_\varepsilon(t,t_0)=0$. Thus, only a photoelastic contribution is present in the signal $\Delta \tilde{r}(t)$, governed by the fraction of $VO_2$ in the metallic phase after PIPT.

It is seen in Fig. S6 that the blue and green curves $\Delta \tilde{r}(t)$ measured for different delays do not show any noticeable difference. This is in full agreement with the known fact that the fraction of $VO_2$ in the metallic phase after PIPT starts to change on a longer time scale, i.e. microseconds [39]. The independence of $\Delta \tilde{r}(t)$ of $t_0$ allows the procedure of extracting $\Delta R(t)$ using Eq.(2) (see the main text) for the experiments when $t>t_0>0$. In this procedure the signal $\Delta \tilde{r}(t)$ is used as a function which describes purely photoelastic contribution after the optical pulse at any time including the case when $t_0>0$.

Eq.(2) may give a noticeable deviation from the actual $\Delta R(t)$ in the temporal interval between $t=t_0>0$ and $t=t_0+t_p$, where $t_p\sim 2h/v_{eff}\sim 50$ ps is the phonon propagation time in the $VO_2$ hillock with the height $h$. In this time interval, phonons are injected into the hillock being in insulating phase (i.e. at $t<t_0$) but their further evolution at $t>t_0$ proceeds after PIPT when a certain fraction of $VO_2$ is in metallic phase and possesses different elastic parameters. We suggest that such transient process may lead to an artifact which shows the small residual contribution to $\Delta R(t)$ when $\Delta r_0(t)$ is changing sign and correspondingly $\bar{\varepsilon}(t_0)\approx 0$. An example for this case at $t_0=70$ is shown in Fig. S7.

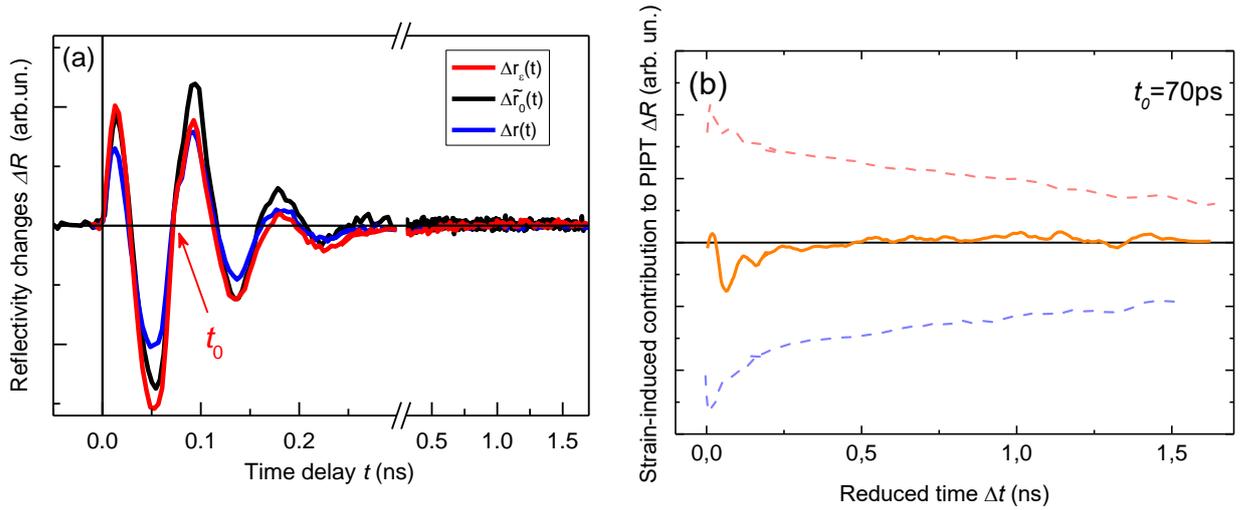

Figure S7. (a) A triad $\Delta r_\varepsilon(t)$, $\Delta r_0(t)$ and $\Delta \tilde{r}(t)$ measured for zero mean strain at the moment $t_0$ of the optical pulse impact: $\bar{\varepsilon}(t_0)=0$. (b) Strain-related contribution to the PIPT $\Delta R(\Delta t - t_0)$ extracted from the data in (a) (solid line). Dashed lines are the two curves $\Delta R(t)$ obtained for $t_0=60$ and 95 ps corresponding to nonzero strain [see Fig 3.(a) in the main text].

The red line in Fig. S7(a) shows the signals $\Delta r_\varepsilon(t)$ in VO$_2$ hillocks under the impact of a picosecond strain pulse measured in the time interval 1700 ps. The black curve is the signal $\Delta r_0(t)$ measured without optical pump. The red arrow indicates the time $t_0=70$ ps when additional optical pump is applied. The value of $t_0$ is tuned so that $\Delta r_0(t_0)=0$. The blue line corresponds to $\Delta \tilde{r}(t)$ whis optical pulse applied before the arrival of the strain pulse to VO$_2$ film, $t_0=-30$ ps. The excitation density $W=8$ mJ/cm$^2$ of the optical pump is above the threshold and below the saturation level for photoinduced phase transition in VO$_2$, $W_T<W<W_S$. Figure S7(b) shows the result of subtraction of photoelastic effect using Eq.(2) of the main text. It is clearly seen that the slow component of calculated $\Delta R(t-t_0)$ is $A_S=0$. There is a nonzero residual contribution to the fast component $A_F$ which could be due to an artifact related to photoelastic contribution subtraction, discussed above. However, some uncertainty in obtaining the evolution for $\Delta R(t-t_0)$ in this short time interval when $\Delta r_0(t_0)=0$ does not change the main conclusions (*i*) and (*ii*) following from the analysis of experimental data in the main text.

### III. Experimental details

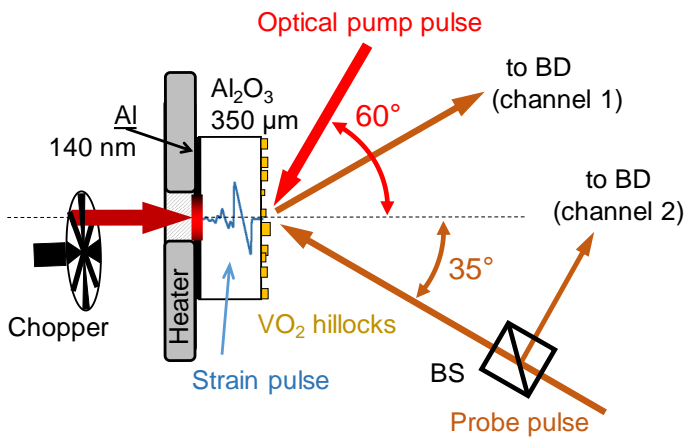

Figure S8. The details of the experimental schematic presented in Fig. 1(d).

The laser source is a 170 fs Yb: KGd(WO$_4$)$_2$ regenerative amplifier with a central photon energy of 1.2 eV and a repetition rate of 5 kHz. The beam from the source is split into three beams (Fig. S8). The *optical pump beam* is incident on VO$_2$ nanohillocks at the angle of 60 deg and its elliptical spot size is of 55×100 μm$^2$. The *second beam which is used to generate the strain pulse* is incident along the normal onto an Al transducer; the spot size is of 110×110 μm$^2$. This beam is modulated with a mechanical chopper at a frequency of $v$=625 Hz. The *probe beam* is, in turn, spit into two by a beamsplitter (BS). One part is incident to the VO$_2$ nanohillocks with the spot size is 25×30 μm$^2$. The reflected probe beam goes onto a channel 1 of the balanced photodetector (BD). The second part of the probe beam is directed straight onto a BD's channel 2 and serves as a reference signal. Use of the balanced detector greatly decreases the laser-source-related noise. Alock-in amplifier, is used to measure the strain-related response, and is locked to the mechanical chopped modulating the beam, exciting strain pulses. The overall sensitivity of the setup reaches $\Delta R/R$~10$^{-6}$. The sample is placed on a copper heater with a good thermal contact, which allows heating the sample from a room temperature 295 K up to 400 K.

### IV. Experimental results at elevated temperature T>T$_c$

In order to check if the strain induced effect described in the main text is indeed a manifestation of the strain-induced modulation of PIPT, we have performed analogous measurements when the sample was heated up to T=360 K, which is reliably above $T_c$ (Fig. S9). We note that no indication of PIPT was found at the pure optical pump-probe experiments at this temperature. As in Fig. 3 in the main text, the black and red curves in Fig, S9 correspond to signals $\Delta r_0(t)$ and $\Delta r(t)$ measured without and with optical pump pulses respectively. The blue curve in (a) corresponds to $\Delta \tilde{r}(t)$, with optical pulse applied before the arrival of the strain pulse to VO$_2$, $t_0$ =-30 ps. The optical pump excitation density $W$=8 mJ/cm$^2$ is above the threshold and below the saturation level for photoinduced phase transition in VO$_2$, $W_T<W<W_S$ obtained at room temperature. The red curves (b, c) correspond to $\Delta r(t)$ when the optical pulse is applied to the VO$_2$ in the presence of a non-zero strain: $\bar{\varepsilon}(t_0)<0$ (b) or $\bar{\varepsilon}(t_0)>0$ (c). As one can see, excitation with the optical pump does not have any effect on the measured reflectivity change when the sample in initially in the metallic state.

### IV.     Effect of strain-induced reflectivity change on PIPT

One can argue that the strain pulses changes the optical properties of the structure under study, including reflectivity at the VO$_2$/Al$_2$O$_3$ and VO$_2$/air interfaces through the photoelastic effect. This, in turn, can alter the optical excitation density in VO$_2$ nanohillocks, and, hence, affect the magnitude of PIPT. From the data in Fig. 1(g) we get that the relative changes of total reflectivity due to photoelastic effect is ~10$^{-4}$, and, hence, the strain-induced changes of the amount light absorbed in VO$_2$ cannot exceed 10$^{-4}$. The changes of the reflectivity due to PIPT [Fig.1(e)] in the absence of strain pulses is ~10$^{-2}$. Therefore, effect on PIPT of strain–induced modulation optical excitation density is estimated to be <~10$^{-6}$. This is two orders of magnitude less than the strain-induced changes of PITP observed in our experiments, and we can confidently rule out such a trivial reason for the strain-induced modulation of PIPT.

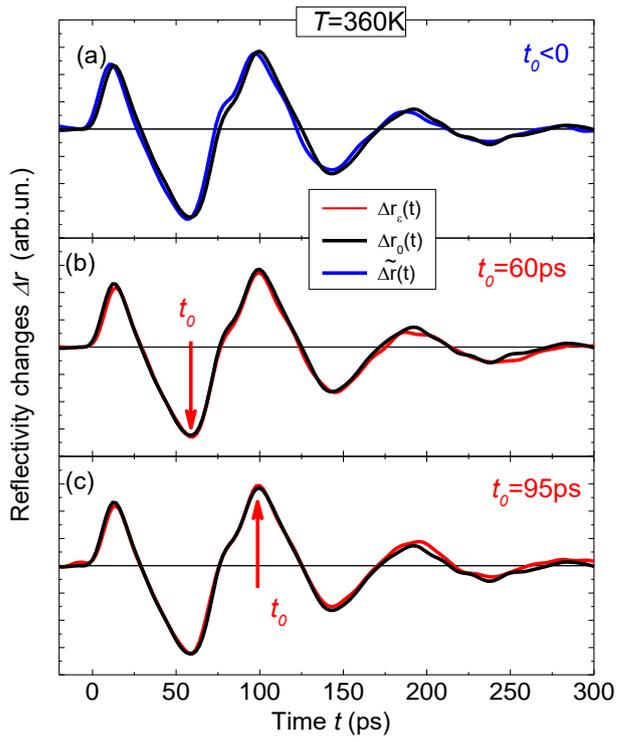

Figure S9. Temporal evolutions of reflectivity changes $\Delta r(t)$ in $VO_2$ hillocks under the impact of picosecond strain pulse at a temperature $T=360$ K which is reliably above the PIPT temperature $T_c=340$ K. Vertical arrows indicate the exact time $t_0$ when the optical pulse is applied.

8